\documentclass[12pt, draftclsnofoot, onecolumn]{IEEEtran}
\usepackage{graphicx}
\usepackage{algorithm}
\usepackage{algorithmicx}
\usepackage[noend]{algpseudocode}
\usepackage{setspace}
\usepackage{cite}
\usepackage{color}
\usepackage{amsmath,amssymb,amsfonts,amsthm}
\usepackage{bm}
\usepackage{enumerate}
\usepackage{multirow}
\usepackage{array, booktabs}
\usepackage{epstopdf}
\usepackage{wasysym}  
\usepackage{textcomp} 
\usepackage[usenames,dvipsnames]{xcolor}
\usepackage{makecell}

\begin{document}
\title{Buffer Occupancy and Delivery Reliability Tradeoffs for Epidemic Routing}

\author{Pin-Yu Chen, Meng-Hsuan Sung, and Shin-Ming~Cheng,~\IEEEmembership{Member,~IEEE}
\thanks{P.-Y. Chen is with the Department of Electrical Engineering and Computer Science, University of Michigan, Ann Arbor, USA. Email: pinyu@umich.edu.}
\thanks{M.-H. Sung and S.-M. Cheng are with the Department of Computer Science and Information Engineering, National Taiwan University of Science and Technology, Taipei, Taiwan. Email: M10215015@mail.ntust.edu.tw and smcheng@mail.ntust.edu.tw.}}

\maketitle

\setstretch{1.9}
\begin{abstract}
      To achieve end-to-end delivery in intermittently connected networks, epidemic routing is proposed for data delivery at the price of excessive buffer occupancy due to its store-and-forward nature. The ultimate goal of epidemic routing protocol design is to reduce system resource usage (e.g., buffer occupancy) while simultaneously providing data delivery with statistical guarantee. Therefore the tradeoffs between buffer occupancy and data delivery reliability are of utmost importance. In this paper we investigate the tradeoffs for two representative schemes: the \textit{global timeout} scheme and the \textit{antipacket dissemination} scheme that are proposed for lossy and lossless data delivery, respectively. For lossy data delivery, we show that with the suggested global timeout value, the per-node buffer occupancy only depends on the maximum tolerable packet loss rate and pairwise meeting rate. For lossless data delivery, we show that the buffer occupancy can be significantly reduced via fully antipacket dissemination. The developed tools therefore offer new insights for epidemic routing protocol designs and performance evaluations.
\end{abstract}


\section{Introduction}
\label{sec_intro}
      Epidemic routing is known to be a promising candidate toward end-to-end data delivery in intermittently connected networks \cite{Vahdat00,Zhang06,Khabbaz12}. Since end-to-end path between the source and the destination nodes might not exist at any one time in such networks, the data are delivered in a store-and-forward fashion, that is, all nodes encountering the source node participate in relaying the data to other nodes until the data are received by the destination node. Although such a data delivery scheme reduces the end-to-end latency and spares the need for routing table updates, it inevitably induces tremendous buffer occupancy for each relaying node. Therefore striking the balance between buffer occupancy and delivery reliability is of utmost importance in epidemic routing protocol design.

      As the data delivery dynamics of store-and-forward routing schemes much resemble the spreads of epidemics \cite{Khelil02,CPY10}, throughout this paper we use the terminology from epidemiology \cite{Hethcote00,Daley01} to model epidemic routing. Analogously, a node is in the infected state if it receives the data and has the ability to deliver the data to surrounding ndoes. A node is in the recovered state if it is immune to the data (i.e., it refuses to receive the data). A node is in the susceptible state if it is neither in the infected state nor in the recovered state (i.e., it will participate in data delivery after receiving the packet). This epidemic model is known as the susceptible-infected-recovered (SIR) model \cite{Hethcote00,Daley01}.


      Due to the spreading nature, before the data reaches the destination node the average number of infected nodes (i.e., the nodes who have received the data) increases monotonically with time. After the destination successfully receives the data, the relaying packets buffered at intermediate nodes become redundant and are expected to be removed. The deletion of packet for a node can be viewed as undergoing the transition from infected state to recovered state, and thus the immunity mechanisms in epidemiology can be applied to resolve excessive \textit{buffer occupancy} problem~\cite{Haas06}. Upon the expiration of the global timer, the nodes carrying the data delete the data from their buffers and therefore the nodes transit from infected state to recovered state, which is analogous to self healing immunity mechanism in epidemiology and is referred to as the \textit{global timeout} scheme in epidemic routing. Furthermore, if the infected nodes delete the data from their buffers and the susceptible nodes declare the data to be obsolete when they update the packet delivery notifications (e.g., ACK sent out by the destination node) with the encountered nodes, such behavior is like vaccinating the susceptible and infected nodes with antidotes against the epidemic, which is referred to as the \textit{antipacket dissemination} scheme.

      In view of the end-to-end data delivery at the transport layer, global timeout scheme is applicable to lossy transmissions where the probability of successful delivery has to be guaranteed, i.e., the packet loss rate is within a tolerable range. On the other hand, antipacket dissemination scheme is suitable for lossless transmissions where all packets need to keep forwarding the data until the reception by the destination node is confirmed.

      Throughout this paper, we investigate the engineering interpretations and the effects of these two immunity schemes, i.e., the buffer occupancy and delivery reliability tradeoffs for epidemic routing. We establish analytical models of the data delivery and buffer occupancy dynamics for both global timeout and antipacket dissemination schemes and specify the utility for epidemic routing. Fog global timeout scheme, we provide a closed form expression for determining the optimal global timeout value such that the packet loss rate is statistically guaranteed to be less than a specified maximum tolerable packet loss rate. Specifically, we prove that with the suggested global timeout value the per-node buffer occupancy depends only on the pair-wise meeting rate and the maximum tolerable packet loss rate and is independent of the number of nodes in the system, indicating the promise of a scalable epidemic routing scheme that strikes the balance between data delivery reliability and buffer occupancy. 

      Regarding antipacket dissemination scheme, we demonstrate the importance of cooperative antipacket dissemination that leads to significant reduction in buffer occupancy. The simulation results show that our models can accurately characterize the data delivery and buffer occupancy dynamics in intermittently connected networks and provide adequate global timeout values to minimize buffer occupancy while constraining packet loss rate. Therefore our models can successfully predict the spatiotemporal data delivery dynamics from a macroscopic view of the entire system and serve as a quick reference for epidemic routing analysis in intermittently connected networks.     
    
      The rest of this paper is organized as follows. Sec. \ref{sec_sys} describes our system model and Sec. \ref{sec_for} formulates the state equations of epidemic routing via the SIR model.  Sec. \ref{sec_tradeoff} specifies the SIR models of the global timeout scheme and the antipacket dissemination scheme  and investigates the buffer occupancy and delivery reliability tradeoffs. The performance evaluation of the tradeoffs between data delivery reliability and buffer occupancy are shown in Sec. \ref{sec_per}. Sec. \ref{sec_related} summarizes the related work for epidemic routing. Finally, Sec. \ref{sec_con} concludes this paper. 

\section{System Model}
\label{sec_sys}

\subsection{Network Model}
      We assume that there are $N$ mobile relaying nodes (including one source node) and one mobile destination node in the network. A node can only transmit packet to another one if both nodes are within transmission range $r$ of one another. For the purpose of analysis, only one packet is to be delivered from the source to the destination and perfect packet reception between two encountered nodes is assumed. The packet delivery delay, denoted by $T_D$, is defined as the 
      duration for transmission from the source to the destination.

      A \textit{store-and-forward} fashion is applied in the packet delivery process, that is, when a node receives a packet, it will store the packet at the buffer and  forward the packet whenever it meets other nodes (i.e., consistently forward the packet to all other nodes within its transmission range $r$). We assume that the inter-meeting time of the intermittently connected nodes is exponentially distributed with mean being the reciprocal of the pairwise meeting rate $\lambda$ \cite{Small03,Haas06,Zhang07}. For analysis purpose we also assume the buffer of every node has infinite size and only one packet is stored at the buffer when a node received duplicated copies. 

\subsection{Mobility Model}
      

      The movement pattern of mobile nodes (i.e., how their velocity and location change over time) is modeled by random waypoint (RWP) and random direction (RD) mobility models described as follows.
      \begin{itemize}
      \item Random Waypoint (RWP) model: Each node randomly and uniformly chooses a point in the specified wrap-around square area as the destination and moves at a constant speed $v$ (uniformly drawn from $[v_{min}, v_{max}]$) toward the point following the shortest distance path. The movement process is repeated once it arrives at the destination point. The pairwise meeting rate $\lambda_{RWP}$ of RWP model is~\cite{Groenevelt05}
      \begin{eqnarray}
      \label{mobility_RWP}
          \lambda_{RWP}=\frac{2\omega rE[V^*]}{L^2},
      \end{eqnarray}
      where $\omega$ is the waypoint constant, $r$ is the transmission radius, $E[V^*]$ is the expected value of relative velocity between two nodes and $L$ is the side length of the area.
      
      \item Random Direction (RD) model: Each node travels in a selected direction $\theta$ (uniformly chosen from $[0, 2\pi]$) for a duration $\tau$ at speed $v$ (uniformly chosen from $[v_{min}, v_{max}]$) in the specified wrap-around square area. The movement process is repeated for each duration. The pairwise meeting rate $\lambda_{RD}$ of RD model is~\cite{Groenevelt05}
      \begin{eqnarray}
      \label{mobility_RD}
          \lambda_{RD}=\frac{2rE[V^*]}{L^2}.
      \end{eqnarray}
      \end{itemize}

\subsection{Immunity Schemes}
      Two immunity schemes for epidemic routing, global timeout and antipacket dissemination schemes,  are illustrated  as follows.
      \begin{itemize}
      \item Global timeout scheme: Fig.~\ref{fig_globaltimeout} describes the process of packet delivery for global timeout scheme at different time instances. A node (the source node) is infected at the initial stage (i.e., at time instance $T_1$). At time instance $T_2$, the packet is delivered from the source node to its encountered node (node 3). The encountered node(s) in the susceptible state store the packet in their buffer and their state changes from susceptible to infected. Then, at time instance $T_3$, the infected nodes continue to carry and deliver the packet to encountered nodes. The process continues until the global timer expires. After the global timer expires (i.e., at time instance $T_g$), all relaying nodes (i.e., nodes which carry the data) delete the data and transit to recovered state.
      \begin{figure}[t]
          \centering
          \includegraphics[width=5in]{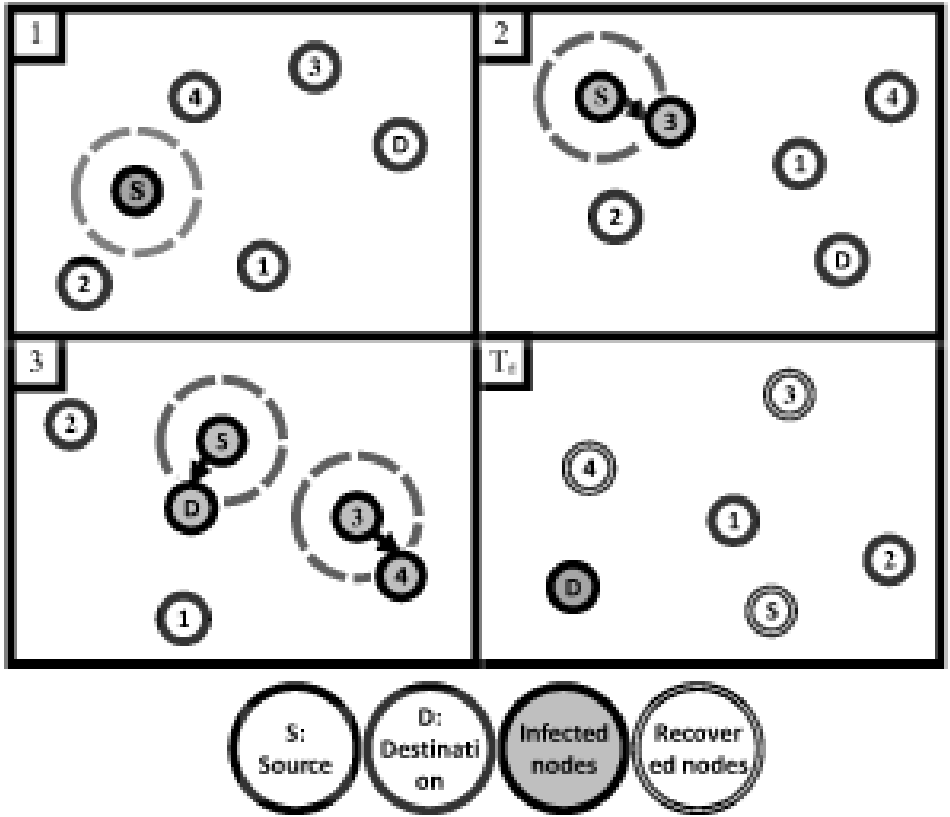}
          \caption{Global timeout scheme. Indices indicate different time instances.}
      \label{fig_globaltimeout}
      \end{figure}
      \begin{figure}[t]
     	\centering
     	\includegraphics[width=5in]{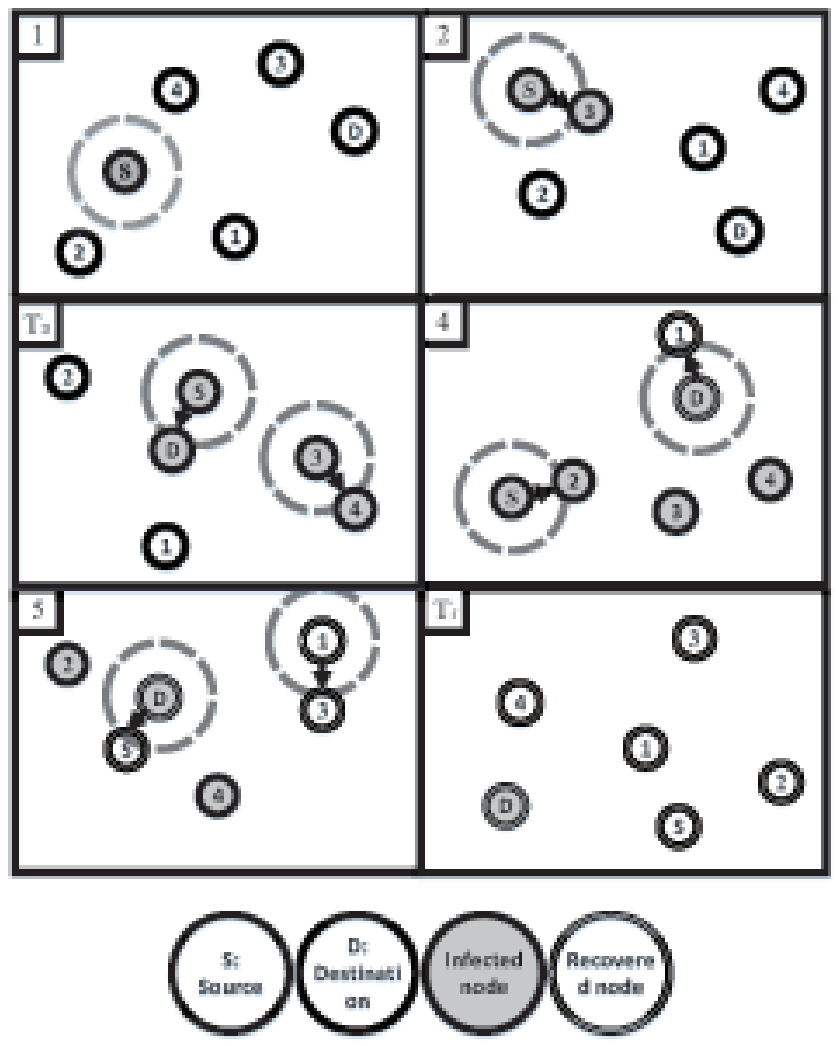}
     	\caption{Antipacket dissemination scheme.}
     	\label{fig_antipacket}
      \end{figure}

      \item Antipacket dissemination scheme: As shown in Fig.~\ref{fig_antipacket}, before the destination node successfully receives the packet (i.e., at time instance $T_D$) the data delivery dynamics of antipackt dissemination scheme are similar to global timeout scheme. After time instance $T_D$ (i.e., time instances $T_4$ and $T_5$), the destination node and the recovered nodes start to deliver the antipacket to encountered nodes. The nodes in the susceptible or infected states transit to recovered state after receiving the antipacket. Upon the reception of antipacket the node in the infected state deletes the packet from its buffer, while the node in the susceptible state declares the packet to be obsolete. However, the infected nodes which have not yet received the antipacket still sustain to deliver packet to encountered nodes. Finally, at the final stage (i.e., time instance $T_f$), the system finishes all transmissions, and there is no infected node in the system.
 
      To investigate the effect of cooperative antipacket dissemination on the system performance, 
      we introduce an antipacket forwarding probability $\kappa$ that governs the willingness to distribute antipacket for each relaying node. As two extreme cases, $\kappa=1$ is the fully antipacket dissemination scenario such that all nodes receiving the antipacket participate in antipacket dissemination. On the other hand, $\kappa=0$ is the null antipacket dissemination scenario such that no node but the destination node is responsible for disseminating the antipacket.    
      \end{itemize}

\subsection{Performance Metrics}
      \begin{itemize}
      \item Buffer occupancy $B$. It is adopted to evaluate the amount of buffer occupied by the packet in the whole network over its end-to-end transmission. In epidemic routing, since nodes will delete duplicated copies, at any time instance the buffer occupancy for a reference packet is the accumulated number of infected nodes in the system. 
      \item Delivery reliability. We adopt the average packet loss rate of several end-to-end data deliveries as the performance metric and compare it with a specified \textit{maximum tolerable packet loss rate} (denoted by $\epsilon$). 
      \end{itemize}

\section{Formulation}
\label{sec_for}
\subsection{SIR model}

      Using SIR model, at any time instance each node is either in the susceptible (S), infected (I), or recovered (R) state. A node which carries the data to be delivered is an infected node, and a node which carries the successful delivery notification (e.g., ACK from the destination node) is a recovered (immune) node. An infected node transits to recovered state upon the global timer expiration or antipacket reception. A susceptible node can either transit to the infected state or recovered state depending on whichever data or successful delivery notification come first. Let $S(t)$, $I(t)$ and $R(t)$ denote the normalized susceptible, infected and recovered population at time $t$, respectively, i.e., $S(t)+I(t)+R(t)=1$. The number of nodes in state $X$ at time $t$ is denoted by $\hat{X}(t)=NX(t)$, where $X=\{S,I,R\}$.

\subsection{Fluid Analysis of SIR model}
      By substituting the relation $S(t)=1-I(t)-R(t)$ and assuming the state equations $X(t)$ to be continuous and nonnegative valued, we have, for a small interval $\Delta t$,           
      \begin{IEEEeqnarray}{rCl}
      \label{eqn_fluid_I}
              I(t+\Delta t)= I(t)+\Upsilon_{S \rightarrow I}(t)\Delta t
              -\Upsilon_{I \rightarrow R}(t) \Delta t,
      \end{IEEEeqnarray}
      where $\Upsilon_{X \rightarrow Y}(t)$ is the expected population transition rate from state $X$ to state $Y$ at time $t$.

      We obtain the first-order ordinary differential equation (ODE) of the infected state equation as
      \begin{align}
      \label{eqn_ODE_I}
          \dot{I}(t)&=\lim_{\Delta t \rightarrow 0}\frac{I(t+\Delta t)-I(t)}{\Delta t}=\Upsilon_{S \rightarrow I}(t)-\Upsilon_{I \rightarrow R}(t)\nonumber\\
          &\triangleq G_I(I(t),R(t)).
      \end{align}

      Similarly, the ODE of recovered state equation is
      \begin{align}
      \label{eqn_ODE_R}
          \dot{R}(t)&=\Upsilon_{I \rightarrow R}(t)+\Upsilon_{S \rightarrow R}(t) \nonumber\\
          &\triangleq G_R(I(t),R(t)).
      \end{align}

      The ODE equations of SIR model evolve with $G_I$ and $G_R$ and these two functions are determined by the routing protocol of our interest, which will be specified in Sec. \ref{sec_tradeoff}.

\subsection{Data Delivery Probability Function and Buffer Occupancy}
      With the exponential pairwise meeting rate $\lambda$, the probability that the destination node receives the data at time $t$ can be evaluated by the probability function $P(t)$, and the state equation of $P(t)$ can be derived as 
      \begin{align}
      \label{eqn_CDF_ODE}
      \dot{P}(t)&=\lim_{\Delta t \rightarrow 0}\frac{P(t+\Delta t)-P(t)}{\Delta t} \nonumber \\
              &=\lim_{\Delta t \rightarrow 0}\frac{\mathbb{P}(T_D>t)-\mathbb{P}(T_D>t+\Delta t)}{\Delta t} \nonumber \\
              &=\lim_{\Delta t \rightarrow 0}\frac{\mathbb{P}(T_D \in(t,t+\Delta t])}{\Delta t} \nonumber \\
              &=\lim_{\Delta t \rightarrow 0}\frac{\mathbb{P}(T_D \in(t,t+\Delta t]|T_D>t)\mathbb{P}(T_D>t)}{\Delta t} \nonumber \\
              &=\lim_{\Delta t \rightarrow 0}\frac{I(t) \lambda \Delta t[1-P(t)]}{\Delta t} \nonumber \\
              &=\lambda I(t) [1-P(t)].
      \end{align}
(\ref{eqn_CDF_ODE}) specifies the rate of increment in $P(t)$ at time $t$, which is associated with the pairwise meeting $\lambda$ and infected population $I(t)$.

      Solving (\ref{eqn_CDF_ODE}) with the initial condition $P(0)=0$, we obtain the analytical expression of $P(t)$ as
      \begin{IEEEeqnarray}{rCl}
      \label{eqn_CDF}
          P(t)=1-\exp\left( - \lambda \int_{0}^{t} I(\tau) d \tau\right).
      \end{IEEEeqnarray}

      To guarantee the delivery reliability for lossy data delivery, it is required that upon the expiration of the global timer $T_g$  the statistical packet loss rate cannot exceed a specified maximum  tolerable packet loss rate $\epsilon$, i.e., 
      $P(T_g) \geq 1-\epsilon$. With (\ref{eqn_CDF}) and the statistical data delivery constraint, we have
      \begin{IEEEeqnarray}{rCl}
      \label{eqn_reliability}
          \int_0^{T_g} I(\tau)d \tau \geq \frac{1}{\lambda} \ln\frac{1}{\epsilon}.
      \end{IEEEeqnarray}

      For buffer occupancy, by Little's formula, the average (system-wise) buffer occupancy for both lossless and lossy data delivery can be evaluated as \cite{Zhang07}
      \begin{IEEEeqnarray}{rCl}
      \label{eqn_buffer}
      B=N \int_{0}^{T_f}I(t)dt,
      \end{IEEEeqnarray}
      which relates to the accumulated infection population from initial time $0$ to the system completion time $T_f$. More precisely, at time $T_f$, the infected population becomes zero either due to global timer expiration or antipacket dissemination such that the data session is complete.

      From (\ref{eqn_CDF}) and (\ref{eqn_buffer}), it is observed that both the delivery reliability and average  buffer occupancy are proportional to the accumulated infected population. Therefore it is of great importance to investigate the tradeoffs between these two metrics for better design of epidemic routing.

\section{Buffer Occupancy and Delivery Reliability Tradeoffs}
\label{sec_tradeoff}
This section specifies the SIR model of the global timeout and antipacket dissemination schemes and investigate the tradeoffs between buffer occupancy and delivery reliability. In particular, we provide an analytical expression of the optimal global timeout value that minimizes buffer occupancy while simultaneously satisfying the statistical delivery reliability constraint.

\subsection{Global Timeout Scheme}
\label{subsec_global}


      In global timeout scheme, the corresponding SIR model can be characterized as
      \begin{IEEEeqnarray}{rCl}
      \label{eqn_SIR_global}
          \left\{
          \begin{array}{ll}
              \dot{I}(t)=\lambda I(t)S(t),&~t \leq T_g, \\
              R(t)=0,&~t \leq T_g, \\
              I(t)=0,&~t > T_g, \\
              R(t)=I(T_g),&~t > T_g, \\
              S(t)+I(t)+R(t)=1,
          \end{array}
          \right.
      \end{IEEEeqnarray}
      where $T_g$ is the global timeout value. The ODE for $I(t)$ is $G_I(t)=\lambda I(t)S(t)$ for $t \leq T_g$, since the data delivery process depends on the coupling of pairwise meeting rate $\lambda$ and how many nodes are infected or can be infected (i.e., susceptible), respectively. Upon the global timer expiration at time $T_g$, all infected nodes discard the data and transit to the recovered state.

      Let $I_0$ be the initially infected population, by (\ref{eqn_SIR_global})
      \begin{IEEEeqnarray}{rCl}
      \label{eqn_I_global}
          I(t)=
          \left\{
          \begin{array}{ll}
              \frac{I_0}{I_0+(1-I_0)\exp\{-\lambda t\}},&~t \leq T_g, \\
              0,&~t > T_g. \\
          \end{array}
          \right.
      \end{IEEEeqnarray}

      From (\ref{eqn_reliability}), given the maximum tolerable packet loss rate $\epsilon$, the optimal global timeout value $T_g^*$ can be obtained by solving
      \begin{IEEEeqnarray}{rCl}
      \label{eqn_global_TTL}
          \int_0^{T_g^*} \frac{I_0}{I_0+(1-I_0)\exp\{-\lambda \tau\}} d \tau = \frac{1}{\lambda} \ln\frac{1}{\epsilon}.
      \end{IEEEeqnarray}

      Since $\int_0^T \frac{1}{1+b\exp\{-a\tau\}} d\tau=\frac{1}{a}\ln\frac{\exp\{aT\}+b}{1+b},~\forall~a,b>0$, we obtain the optimal global timeout value
      \begin{IEEEeqnarray}{rCl}
      \label{eqn_global_TTL_optimal}
          T_g^*=\frac{1}{\lambda} \ln\left[\left(1+\frac{1-I_0}{I_0}\right)\epsilon^{-1}-\frac{1-I_0}{I_0}\right].
      \end{IEEEeqnarray}

      Note that $T_g^*\rightarrow 0$ as $\lambda \rightarrow \infty$, suggesting that data delivery benefits from frequent encounters.
      Moreover, from (\ref{eqn_global_TTL_optimal}), if $I_0=O(\frac{1}{N})$, then $T_g^*=O(\ln(N))$. This suggests that $T_g^*$ scales logarithmically with $N$ when $\epsilon$ and $\lambda$ are fixed. Since $T_f \geq T_g$, from (\ref{eqn_buffer}) and (\ref{eqn_SIR_global}), the traffic and reliability tradeoffs can be represented by the Pareto contour
      \begin{IEEEeqnarray}{rCl}
      \label{eqn_tradeoff}
      	B^*=\frac{N}{\lambda}\ln{\frac{1}{\epsilon}}.
      \end{IEEEeqnarray}

      The Pareto contour suggests that, with proper selection of the global timeout value $T_g^*$ in (\ref{eqn_global_TTL_optimal}), the optimal (minimum) average (system-wise) buffer occupancy $B^*$  depends on the population size $N$, the pairwise meeting rate $\lambda$, and the maximum tolerable packet loss rate $\epsilon$. It is easy to see that frequent encounters (large $\lambda$) or loose statistical delivery constraint (large $\epsilon$) can lead to small buffer occupancy, and vice versa. Moreover, from (\ref{eqn_tradeoff}) the per-node optimal buffer occupancy is $\frac{1}{\lambda}\ln{\frac{1}{\epsilon}}$, which does not depend on the number of nodes in the network. This suggests that with proper selection of the global timeout value  $T_g^*$ in (\ref{eqn_global_TTL_optimal}), the global timeout scheme can be scalable for epidemic routing.

\subsection{Antipacket Dissemination Scheme}
\label{subsec_antipacket}

      The SIR model for the antipacket dissemination scheme can be characterized as
      \begin{IEEEeqnarray}{rCl}
      \label{eqn_SIR_anti}
      \left\{
            \begin{array}{ll}
                \dot{I}(t)=\lambda I(t)S(t),&~t < T_D, \\
                \dot{I}(t)=\lambda I(t)S(t)-\lambda \kappa R(t)I(t)-\frac{\lambda}{N}I(t),&~t \geq T_D, \\
                R(t)=0,&~t < T_D, \\
                \dot{R}(t)=\left[\lambda \kappa R(t)+\frac{\lambda}{N}\right]\left[I(t)+S(t)\right],&~t \geq T_D, \\
                S(t)+I(t)+R(t)=1,
            \end{array}
            \right.
      \end{IEEEeqnarray}
      where $T_D$ is the time instance that the destination received the data. The $\frac{\lambda}{N}$ term represents the meeting rate of a node encountering the destination node. The ODE equation for $R(t)$ is $G_R(t)=\left[\lambda \kappa R(t)+\frac{\lambda}{N}\right]\left[I(t)+S(t)\right]$ for $t \geq T_D$ since nodes in the infected and susceptible states will transit to the recovered state with probability $\kappa$ once they encountered a recovered node or the destination node. Similarly, the ODE equation for $I(t)$ depends on the coupling of $I(t)S(t)$ and $R(t)I(t)$ due to the antipacket dissemination scheme.

      Following (\ref{eqn_SIR_anti}), we obtain
      \begin{IEEEeqnarray}{rCl}
      \label{eqn_SIR_anti2}
          \left\{
          \begin{array}{ll}
              I(t)=\frac{I_0}{I_0+(1-I_0)\exp\{-\lambda t\}},&~t < T_D,\\
              I(t)=1-R(t)-S(t),&~t \geq T_D.
          \end{array}
          \right.
      \end{IEEEeqnarray}

      Due to the fact that $R(T_D)=1/N$ (i.e., one node encountered the destination at time $T_D$), neglecting the term we have for $\kappa>0$,     
      \begin{align}
      \label{eqn_SIR_anti_R}
          R(t)=\frac{1}{1+(N-1)\exp\{-\lambda \kappa(t-T_D)\}},&~t \geq T_D.
      \end{align}     
      If $\kappa=0$, we have     
      \begin{align}
      \label{eqn_SIR_anti_R2}
          R(t)=1-\frac{N-1}{N}\exp\{-\frac{\lambda}{N}(t-T_D)\},&~t \geq T_D.
      \end{align}
      Moreover, by neglecting the $\frac{\lambda}{N}$ term in (\ref{eqn_SIR_anti}), we have
      \begin{align}
      \label{eqn_SIR_anti_S_eqn}
          \dot{S}(t)=-\lambda S(t) \left[ I(t)+\kappa R(t)\right],~t \geq T_D.
      \end{align}
      For two extreme cases ($\kappa=1$ or $\kappa=0$), we have
      \begin{align}
      \label{eqn_SIR_anti_S}
          S(t)=\frac{1-I_0}{I_0+(1-I_0)\exp\{\lambda t\}}.
      \end{align}

      Let 
      \begin{IEEEeqnarray}{rCl}
      \left\{
      \begin{aligned} %
      g(T_D)&=N \int_{0}^{T_D} I(t) dt \nonumber \\ &=N\int_{0}^{T_D} \frac{I_0}{I_0+(1-I_0)\exp\{-\lambda t\}} dt\\ &= N \ln (I_0\exp\{\lambda T_D\}+1-I_0), \\
      h(T_D)&=N \int_{T_D}^{T_f} S(t) dt \nonumber \\ &= N\frac{1-I_0}{I_0} \ln \frac{I_0 \exp \{ -\lambda T_D\}+1-I_0} {I_0 \exp\{-\lambda T_f\}+1-I_0}, \\
      f_0(T_D)&=N\int_{T_D}^{T_f} \frac{N-1}{N}\exp\{-\frac{\lambda}{N}(t-T_D)\} dt\\ &=N(N-1)[ 1-\exp\{-\frac{\lambda}{N}(T_f-T_D)\} ] \\
      f_{\kappa}(T_D)&=N \int_{T_D}^{T_f} R(t)dt \nonumber \\ &=\frac{N}{\kappa} \ln \frac{\exp \{\lambda \kappa (T_f-T_D)\}+N-1}{N}.
      \end{aligned}
      \right.
      \end{IEEEeqnarray}

      Since $T_f \geq T_D$, the buffer occupancy becomes
      \begin{align}
      \label{eqn_buffer_early}
          B&=N \left(  \int_{0}^{T_D} I(t) dt+\int_{T_D}^{T_f} \left[1-R(t)-S(t)\right] dt \right) \\
          &=\left\{
          \begin{array}{ll}
              g(T_D)-h(T_D)+f_0(T_D),&\kappa=0,  \\
              g(T_D)-h(T_D)+N(T_f-T_D)-f_{\kappa}(T_D), &\kappa \in (0,1],\nonumber
          \end{array}
          \right.
      \end{align}
      where $T_f=\{\min~t >0 : I(t)=0\}$.

      In general, the buffer occupancy caused by the antipackets is not a major concern since the size of antipacket is negligible compared with that of data packet. However, if the buffer occupancy of the antipackets may affect the system performance, we can apply the global timeout scheme to eliminate the obsolete antipackets as proposed in \cite{Zhang07}.

\section{Numerical Results}
\label{sec_per}
      This section conducts extensive simulation experiments to validate the analytical model and the utility of the global timeout and antipacket dissemination schemes. We use the setting that there are $N$ moving nodes in a  wrap-around square area with side length $L$ and we randomly select a source-destination pair for end-to-end transmission with $I_0=1/N$. We adopt RWP and RD mobility models in the simulation. In both mobility models, the nodal moving speed is independently and uniformly drawn from $v_{min}=4~km/h$ to $v_{max}=10~km/h$. The transmission range of each node is set to be $r=0.1~km$. Following the parameter setup in~\cite{Groenevelt05}, the expected relative velocity $E[V^*]$ is $8.7~km/h$ for RWP and $9.2~km/h$ for RD, respectively. For RWP, the RWP constant $\omega$ is $1.3683$. From (\ref{mobility_RWP}) and (\ref{mobility_RD}), we know that $\lambda$ and $L$ have one-to-one mapping when the values $r$, $E[V^*]$, $N$ and $\omega$ are fixed. Following the suggestions in \cite{Groenevelt05}, we investigate the cases when the pairwise meeting rates are $0.14817$ and $0.37043$, where the corresponding side lengths are $2.5352~km$ and $4~km$, respectively. The system completion time is set to be $T_f = 20000$ seconds. Two quality-of-service (QoS) requirements corresponding to lossy and lossless data transmissions are considered from the aspect of maximum packet loss rate $\epsilon$ as follows. 
      \begin{itemize}
      \item lossy data transmission: the packet loss rate is within a tolerable range, i.e., $\epsilon$ is set to be a tolerable small value.
      \item lossless data transmission: no packet loss is allowed, i.e., $\epsilon=0$.
      \end{itemize}

\subsection{Global Timeout Scheme}
      Obviously, the setup of the global timer affects the packet loss rate since if the global timer expires before the time instance that the destination receives the packet, the packet reception can not be successful. To control the packet loss rate, a reasonable global timer (such as (\ref{eqn_global_TTL_optimal})) shall be determined, which is further associated with the corresponding buffer occupancy. The relationship among global timer, packet loss rate, buffer occupancy are investigated via simulations in this subsection.

\subsubsection{Relationship between optimal global timer $T^*_g$ and maximum packet loss rate $\epsilon$}

      \begin{figure}[t]
          \centering
          \includegraphics[width=5in]{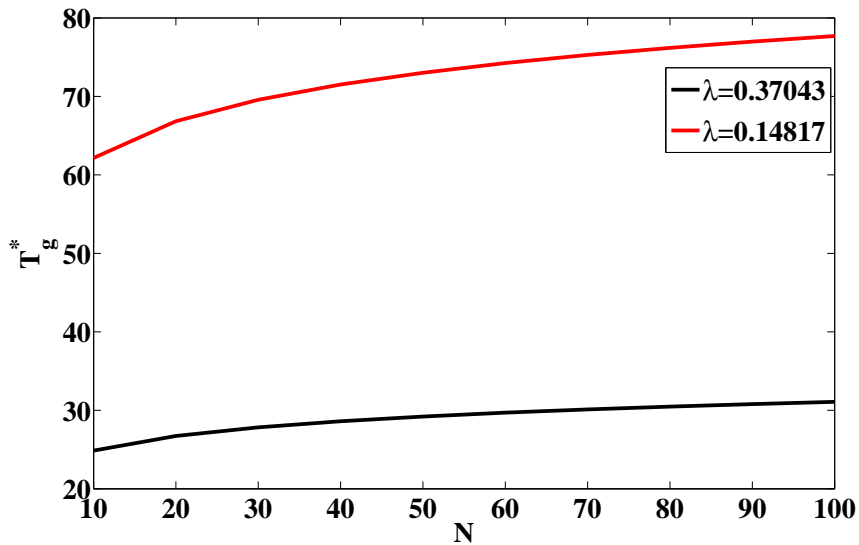}
          \caption {Optimal global timeout value with respect to various $N$ and $\lambda$. The system parameters are set as $\epsilon=10^{-3}$, $I_0=1/N$, $T_f=20000$, $r=0.1~km$. For different $N$, $L$ ranges from $0.8~km$ to $2.5352~km$ when $\lambda=0.37043$ and ranges from $1.2651~km$ to $4~km$ when $\lambda=0.14817$. RWP mobility model is applied. $T_g^* \rightarrow 0$ when $\lambda \rightarrow \infty$ suggests that data delivery benefits from frequent encounters (large $\lambda$). When $I_0=O(\frac{1}{N})$, $T_g^*$ scales logarithmically with $N$ as predicted by (\ref{eqn_global_TTL_optimal}).}
      \label{fig_fixTG_optimalTg}
      \end{figure}

      \begin{LaTeXdescription}
      \item [Effects of $N$ and $\lambda$ on $T^*_g$.]
      For global timeout scheme, the optimal global timeout value with respect to the total population $N$ obtained via equation (\ref{eqn_global_TTL_optimal}) by a given maximum tolerable packet loss rate $\epsilon = 10^{-3}$ is shown in Fig.~\ref{fig_fixTG_optimalTg}. To make a fair comparison among the cases of different $N$, we fix the pairwise meeting rate $\lambda$ by adjusting the moving speed, that is, when the number of users is larger, everyone shall move slower. As a result, in the case of $N$ ranges from $10$ to $100$, the corresponding $L$ is ranges from $0.8~km$ to $2.5352~km$ when $\lambda=0.37043$ and ranges from $1.2651~km$ to $4~km$ when $\lambda=0.14817$.

      We can observe from Fig.~\ref{fig_fixTG_optimalTg} that given the maximum tolerable packet loss rate, the optimal global timeout value $T_g$ decreases if the pairwise meeting rate $\lambda$ increases. It is due to the reason that the packet is expected to be delivered with a faster speed when $\lambda$ is larger, and thus the destination will receive the packet earlier. As expected from (\ref{eqn_global_TTL_optimal}), when $I_0=O(\frac{1}{N})$, the optimal global timeout value increases logarithmically with $N$. The reason behind this phenomenon is that as $N$ increases, to maintain the same $\lambda$, $V$ will decrease, which implies that users move slower. As a result, the packet propagation speed decreases and the destination will receive the packet later.

      \begin{figure}[t]
          \centering
          \includegraphics[width=5in]{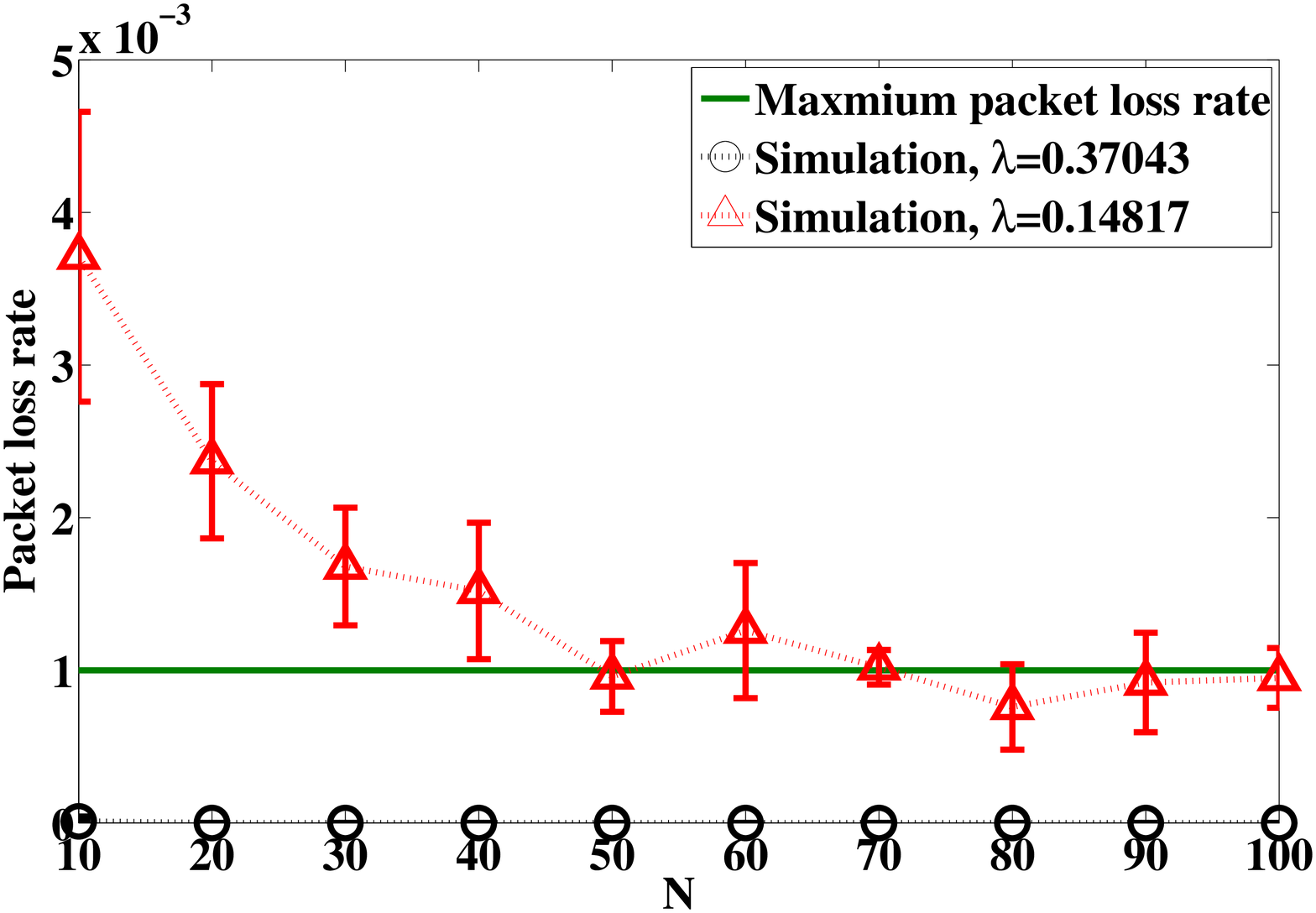}
          \caption {Packet loss rate with respect to various $N$ and $\lambda$ under a fixed $\epsilon=10^{-3}$. The system parameters are the same as that in Fig.~\ref{fig_fixTG_optimalTg}.}
      \label{fig_fixTG_epsilon}
      \end{figure} 

      Fig.~\ref{fig_fixTG_epsilon} depicts the effects of $N$ and $\lambda$ on packet loss rate under fixed $\epsilon$ via simulation experiments following the same parameter setup in Fig.~\ref{fig_fixTG_optimalTg}. In particular, the suggested optimal global timer derived from (\ref{eqn_global_TTL_optimal}) is applied in the simulation experiment to investigate the resulting packet loss rate. This figure shows that the packet loss rate is around the desired value $10^{-3}$ for different population $N$. For small $N$ the simulation results may deviate from the desired packet loss rate due to large deviation of mean-field approximation to SIR model. The asymptotic result in (\ref{eqn_global_TTL_optimal}) shows that optimal global timeout $T_g^* \rightarrow 0$ as pairwise meeting rate $\lambda \rightarrow \infty$, suggesting that the global timeout value $T_g$ can be made arbitrarily small if the pairwise meeting rate approaches infinity. 
      \begin{figure}[t]
          \centering
          \includegraphics[width=5in]{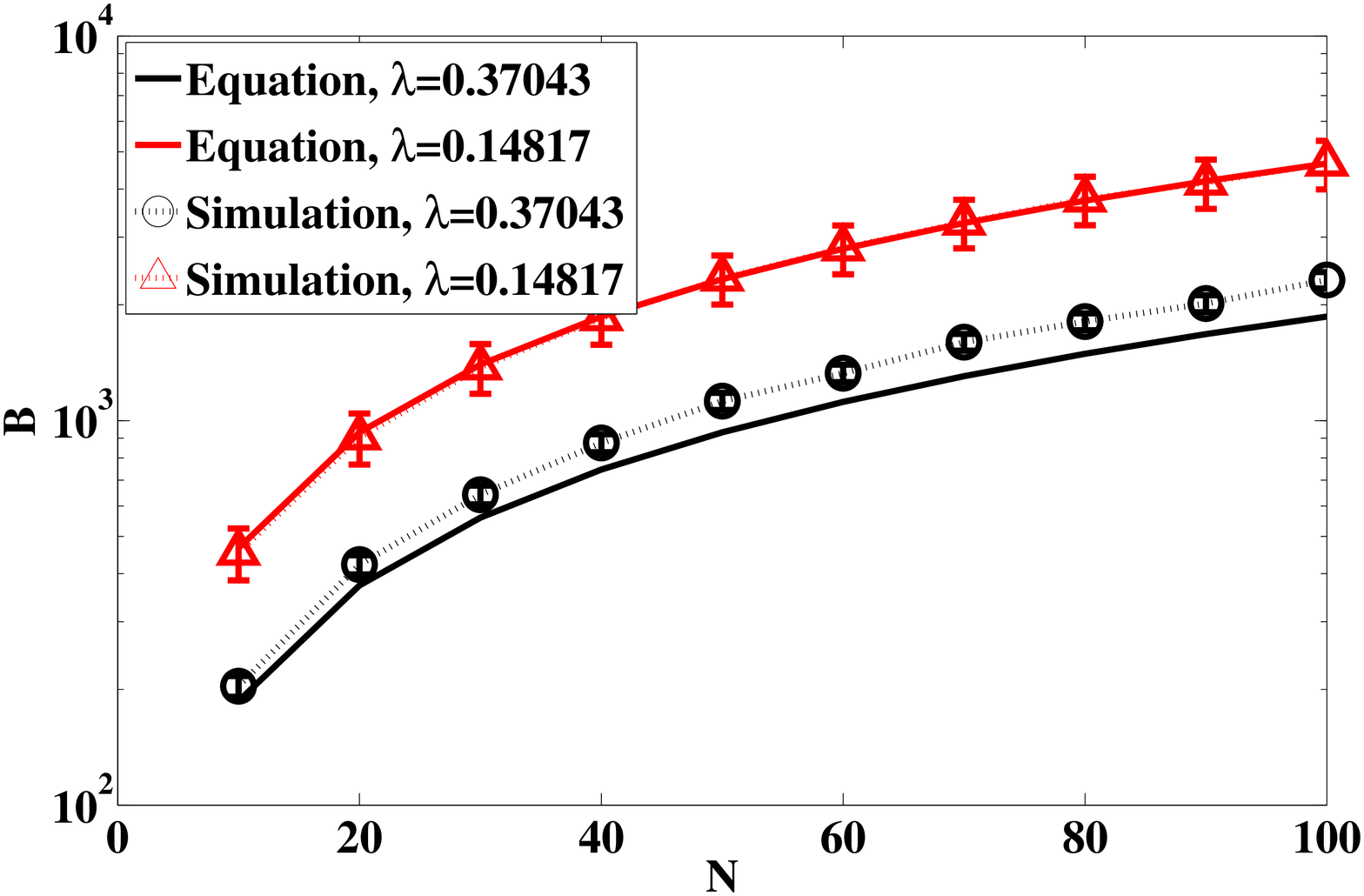}
          \caption {Buffer occupancy with respect to various $N$ and $\lambda$ under a fixed $\epsilon$. The system parameters are same as that in Fig.~\ref{fig_fixTG_optimalTg}.}
      \label{fig_fixTG_buffer}
      \end{figure}

      \item [Effects of $N$ and $\lambda$ on $B$.] Adopting the optimal global timeout value $T_g^*$ in Fig.~\ref{fig_fixTG_optimalTg}, Fig.~\ref{fig_fixTG_buffer} depicts the effects of $N$ and $\lambda$ on $B$ under a fixed $\epsilon$, including both analytical and simulation results. We can observe that the correctness of the analytical model in (\ref{eqn_tradeoff}) is verified by the simulation experiments. This figure also shows that with proper selection of the optimal global timeout value $T^*_g$, the optimal (minimum) buffer occupancy $B$ increases linearly with the population $N$. It is due to the fact that when $N$ becomes larger, the time that the destination receives the packet becomes later, and thus the buffer occupancy $B$ becomes larger. The reason why smaller $\lambda$ incurs larger $B$ is similar. 
      \end{LaTeXdescription}

\subsubsection{Tradeoff between maximum packet loss rate $\epsilon$ and buffer occupancy $B$}

      \begin{figure}[t]
          \centering
          \includegraphics[width=5in]{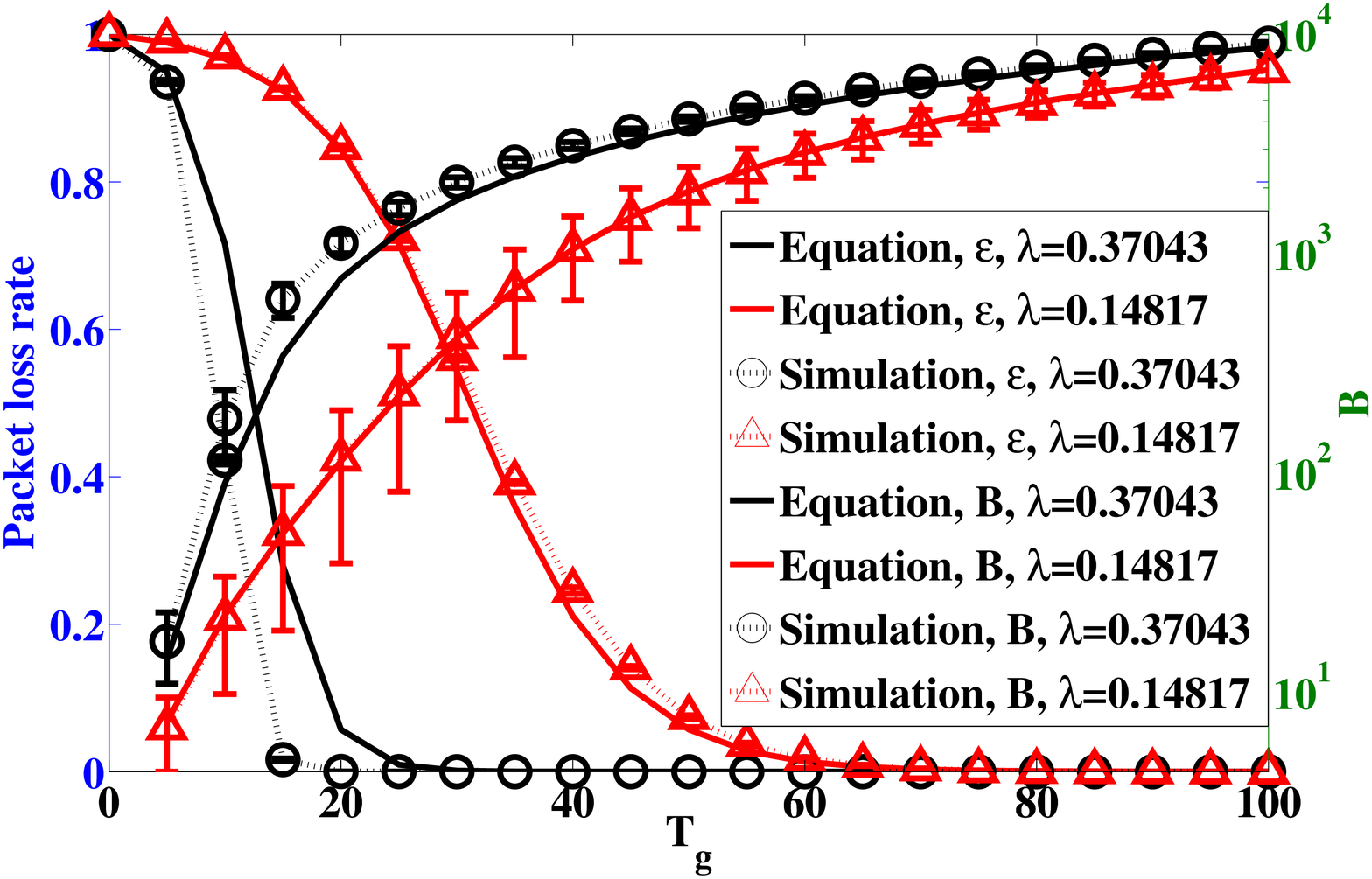}
          \caption { Packet loss rate and buffer occupancy with respect to $T_g$. The system parameters are set as $N=100$, $I_0=1/N$, $T_f=20000$, $r=0.1~km$, $L=2.5352~km$ when $\lambda=0.37043$ and $L=4~km$ when $\lambda=0.14817$. RWP mobility model is applied.
           Increasing $T_g$ leads to the decrease in packet loss rate decreases and the increase in buffer occupancy, as predicted by our 
           analysis from (\ref{eqn_global_TTL}) and (\ref{eqn_buffer}), respectively.}
      \label{fig_TgDifferent_TgepsiBuffer}
      \end{figure}

      \begin{LaTeXdescription}
      \item [Effects of $T_g$ and $\lambda$ on packet loss rate and $B$.] Fig.~\ref{fig_TgDifferent_TgepsiBuffer} depicts the effects of $T_g$ and $\lambda$ on both packet loss rate and buffer occupancy $B$. The packet loss rate decreases when global timeout $T_g$ increases since the destination node has more chance to receive the packet. Moreover, when $\lambda$ is larger, packet loss rate is smaller since nodes have more chance to meet each other, which facilitate the packet propagation process. 
      
      Regarding the buffer occupancy $B$, we found in this figure that $B$ becomes larger when $\lambda$ becomes larger or $T_g$ becomes larger. The reason is that in either case, larger number of users will involve in the packet spreading process and more infected population is expected, thereby making $B$ larger. 

      \begin{figure}[t]
          \centering
          \includegraphics[width=5in]{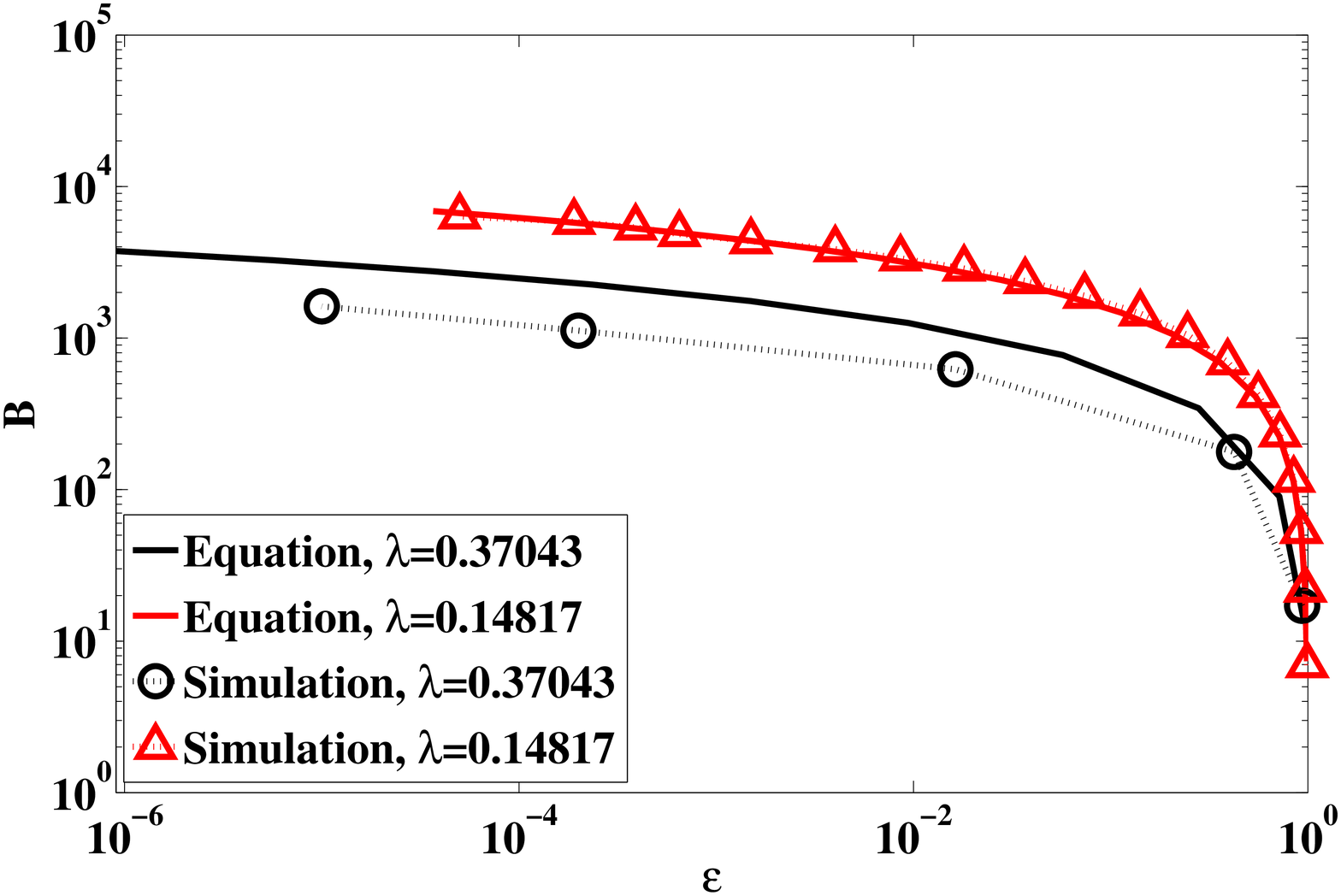}
          \caption {Effects of QoS requirement on buffer occupancy $B$. The system parameters are same as that in Fig.~\ref{fig_TgDifferent_TgepsiBuffer}. $B\rightarrow \infty$ as $\epsilon \rightarrow 0$ implies that the global timeout scheme is inadequate for lossless data delivery at the cost of excessive buffer occupancy. The trends of the change in buffer occupancy is successfully captured by the derivation in (\ref{eqn_tradeoff}).}
      \label{fig_TgDifferent_epsi_Buffer}
      \end{figure}

      Fig.~\ref{fig_TgDifferent_epsi_Buffer} depicts the effects of QoS requirement (i.e., maximum allowable packet loss rate $\epsilon$) on the buffer occupancy $B$. We observe that the tradeoff between $\epsilon$ and $B$ in (\ref{eqn_tradeoff}) is consistent with the simulation results. This figure suggests that the global timeout scheme is inadequate for lossless data delivery at the cost of excessive buffer occupancy since $B \rightarrow \infty$ when maximum tolerable packet loss rate $\epsilon \rightarrow 0$.
      \end{LaTeXdescription}

\subsubsection{Comparisons of Different Mobility Models}
      In this subsection we discuss the effect of mobility models on the epidemic routing with global timeout scheme. Since the correctness of analytical model is validated in the previous subsection, we omit analytical results for the following simulations.


      \begin{figure}[t]
          \centering
          \includegraphics[width=5in]{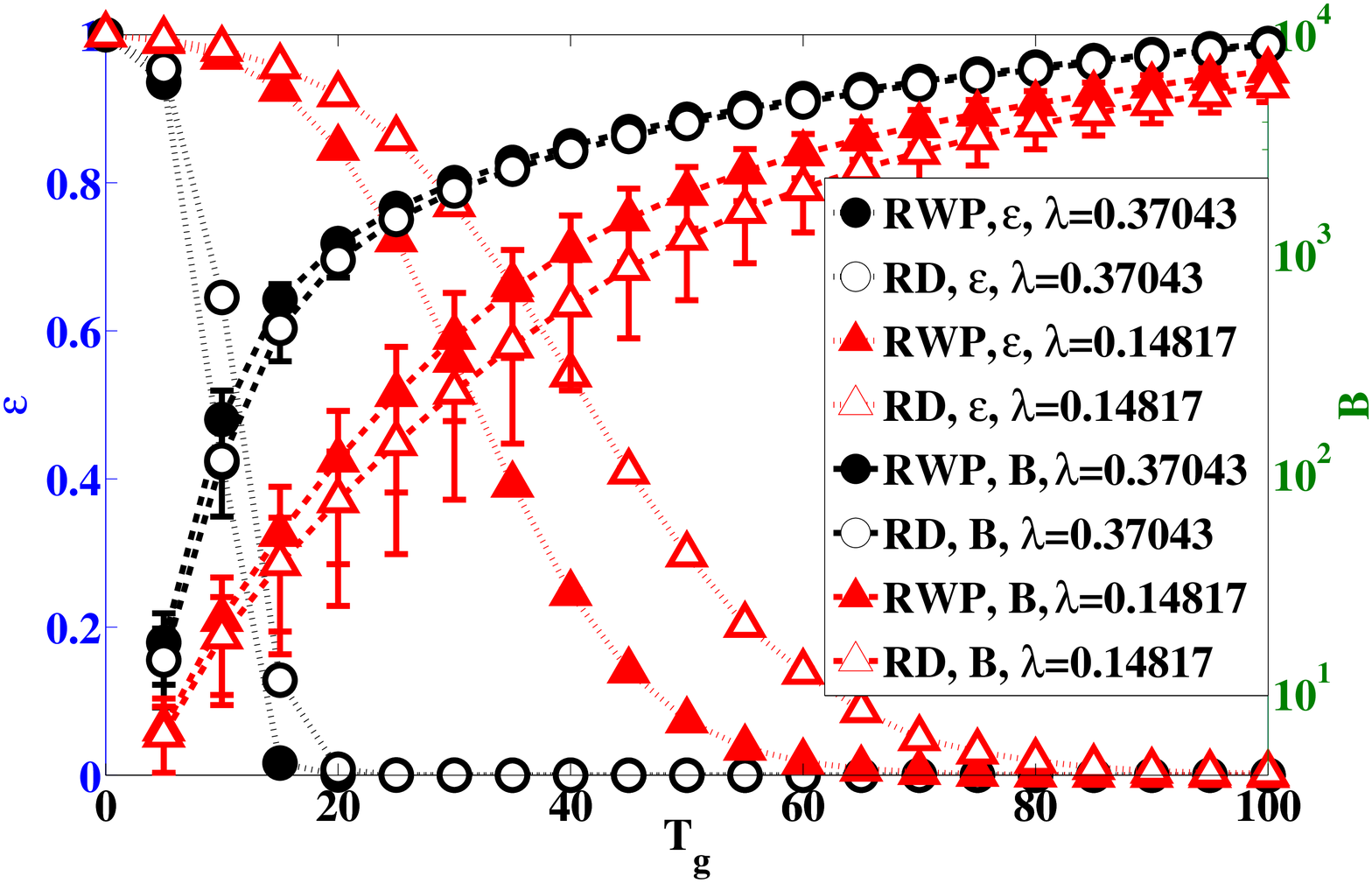}
          \caption{Packet loss rate and buffer occupancy with respect to $T_g$ with both RD and RWP mobility models. The system parameters are set as $N=100$, $I_0=1/N$, $T_f=20000$, $r=0.1~km$, $L=2.5352~km$ when $\lambda=0.37043$ and $L=4~km$ when $\lambda=0.14817$. The speed is generated within $4~km/h$ to $10~km/h$ and corresponding to the expected relative velocity is $8.7~km/h$ for RWP and $9.2~km/h$ for RD. The results show that RWP has smaller packet loss rate than RD and higher buffer occupancy than RD under the same $T_g$.}
      \label{fig_Mobility_TG}
      \end{figure}

      \begin{LaTeXdescription}      
      \item[Effects of mobility model on packet loss rate and $B$.]
      The effects of $\lambda$ and $T_g$ on $\epsilon$ and $B$ with RD and RWP mobility models are illustrated in Fig.~\ref{fig_Mobility_TG}. Obviously, the performance of epidemic routing varies with different mobility models. However, tradeoffs between packet loss rate and buffer occupancy in both models follow the same trend. The RWP has better reliability than RD under the same $T_g$, which is similar to the findings in~\cite{Camp02}. It leads an important result that the packet loss rates of both RD and RWP are smaller than the given QoS constraint $\epsilon$ under the optimal timer $T^*_g$ derived by our analytical model.
      \end{LaTeXdescription}

\subsection{Antipacket Dissemination Scheme}
      In this subsection, we evaluate the performance of epidemic routing with antipacket dissemination scheme from the aspect of how the spreading of antipackets assists in the reduction of buffer occupancy. The forwarding probability of antipacket at each node (i.e., $\kappa$) is introduced, where $\kappa=0$ and $\kappa=1$ respectively represent the null and the fully antipacket dissemination scenarios. Comparing with the fully antipacket dissemination where all nodes participate in antipacket spreading, in null antipacket dissemination, only destination node spreads antipacket when it meets other nodes.

\subsubsection{Relationship between buffer occupancy $B$ and delay $T_D$.}
      \begin{LaTeXdescription}



      \begin{figure}[t]
          \centering
          \includegraphics[width=5in]{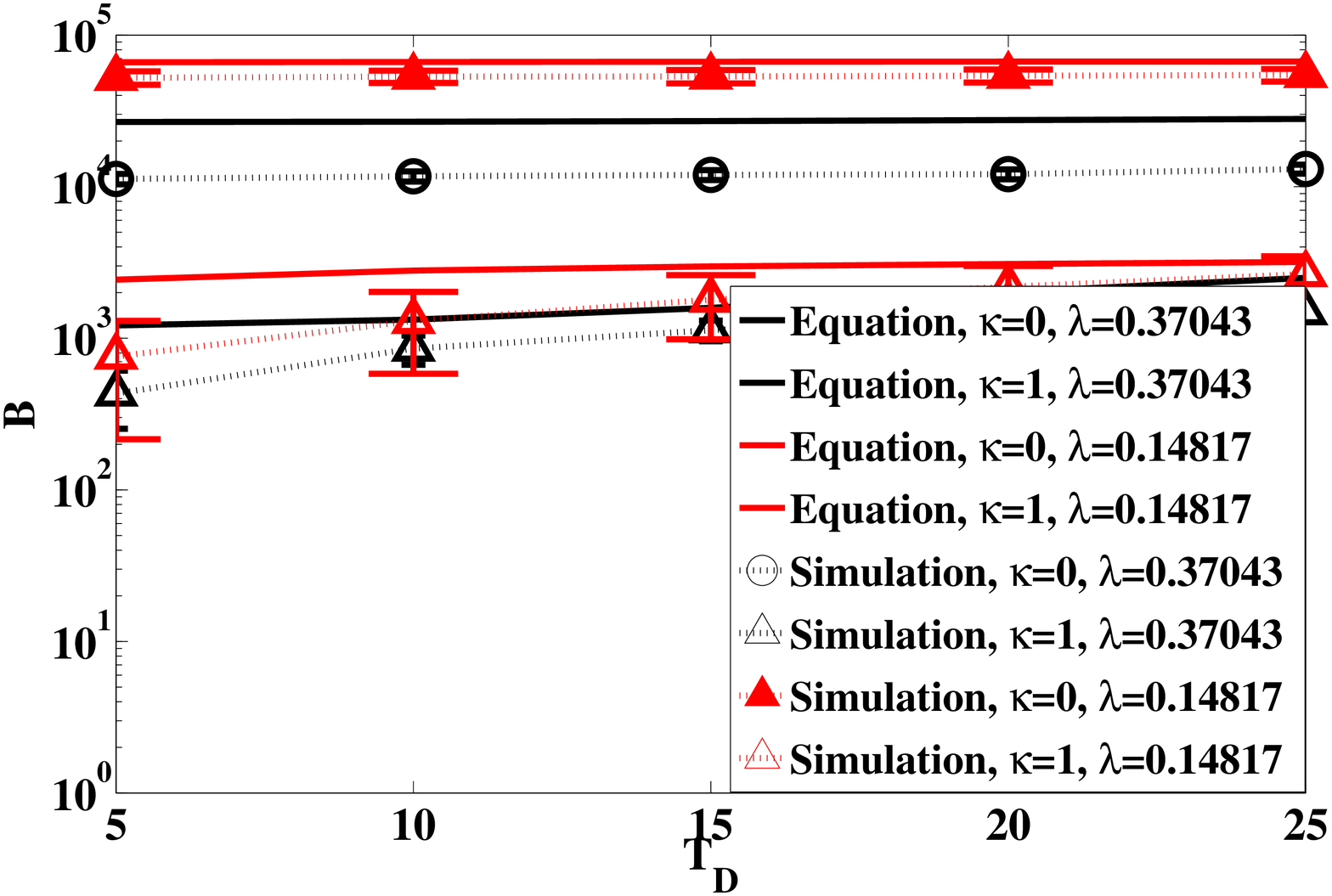}
          \caption{Buffer occupancy with respect to the delivery delay. The system parameters are set as $N=100$, $I_0=1/N$, $T_f=20000$, $r=0.1~km$, $L=2.5352~km$, $L=2.5352~km$ when $\lambda=0.37043$ and $L=4~km$ when $\lambda=0.14817$. RWP mobility model is applied. The buffer occupancy when $\kappa=1$ is significantly smaller than that when $\kappa=0$. Furthermore, the simulated buffer occupancy is shown to be consistent with the analysis in (\ref{eqn_buffer_early}).}
      \label{fig_anti_k0k1}
      \end{figure}

      \item[Effects of $T_D$, $\lambda$, $\kappa$ on $B$.]
      For antipacket dissemination scheme, we can obtain predicted buffer occupancy $B$ from (\ref{eqn_buffer_early}). Figs.~\ref{fig_anti_k0k1} depict the effects of $T_D$, $\lambda$, and $\kappa$ on buffer occupancy $B$. This figure shows that as pairwise meeting rate $\lambda$ becomes larger, buffer occupancy $B$ becomes smaller. When pairwise meeting rate increases, both packet spreading and packet dissemination are facilitated, where the former one incurs buffer occupancy while the latter one alleviates the buffer occupancy. This figure therefore told us that the benefits of antipacket cover the costs from packet spreading. 

      We also observe that $B$ in fully antipacket dissemination scheme ($\kappa=1$) is smaller than that in null antipacket dissemination scheme ($\kappa=0$). It is due to the reason that in fully antipacket dissemination scheme, all nodes who has received the antipacket will participate in the antipacket spreading process, which further decreases the number of infected nodes, thereby reducing $B$. 

      Another observed phenomenon is that when $\kappa = 1$, as delivery delay $T_D$ increases, $B$ increases. It is due to the fact that as $T_D$ becomes larger (i.e., it takes more time to deliver the packet to the destination), the antipacket dissemination process will be activated later. In this case, the number of users receiving antipacket becomes smaller and thus the buffer occupancy becomes larger. However, when $\kappa=0$, $B$ increases slightly as $T_D$ increases. This is due to the reason that no matter when the antipacket spreading process is activated, only the destination participates in the antipacket spreading process. As a result, only a few nodes are affected by the process and the improvement of $B$ is negligible. The simulation results related to the buffer occupancy under $\kappa=0$ and $\kappa=1$ verify the correctness of analytical results from (\ref{eqn_buffer_early}). 
      \end{LaTeXdescription}

\subsubsection{Antipacket Forwarding Probability}
      \begin{figure}[t]
          \centering
          \includegraphics[width=5in]{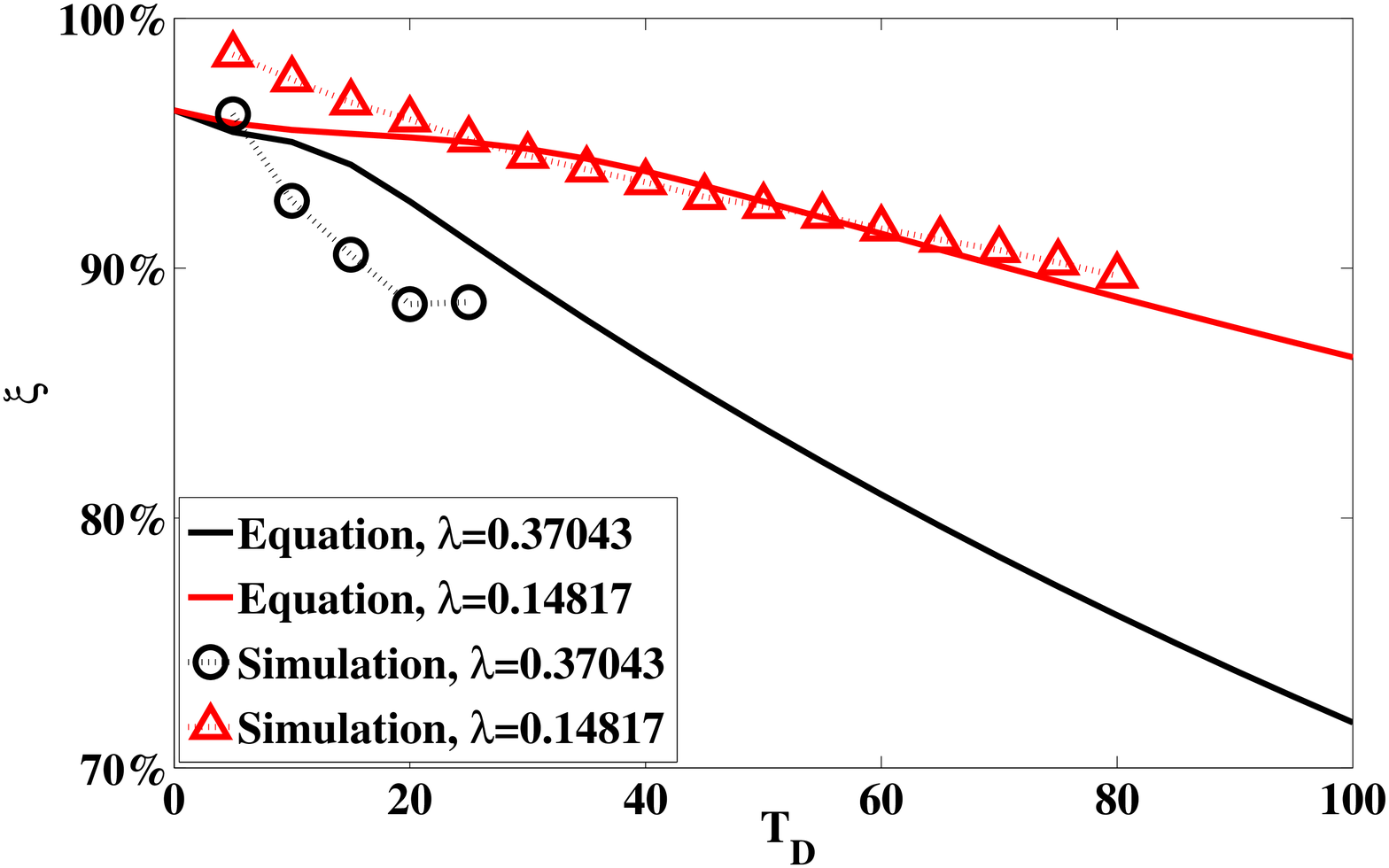}
          \caption{Relative improvement of buffer occupancy with respect to the delivery delay. The system parameters are same as that in Fig.~\ref{fig_TgDifferent_TgepsiBuffer}. Significant buffer occupancy reduction is observed via fully antipacket dissemination scheme. Moreover, when $\lambda$ becomes smaller, the relative improvement becomes larger. The analytical result is obtained from (\ref{eqn_buffer_early}).}
      \label{fig_anti_relative}
      \end{figure}

      To compare the buffer occupancy of different antipacket forwarding probabilities, we define $\xi=\frac{B_{null}-B_{fully}}{B_{null}}$ as the relative improvement of buffer occupancy $B$ in the fully antipacket dissemination from that in the null antipacket dissemination. Obviously, $\xi$ depends on the values of recovered population $R(t)$ in fully and null antipacket dissemination schemes. As a result, $\xi$ can be interpreted as the improvement on buffer occupancy due to the assistance from nodes who participate in antipacket dissemination (except the destination).

      \begin{LaTeXdescription}
      \item[Effects of $T_D$ and $\lambda$ on $\xi$.] Fig.~\ref{fig_anti_relative} plots $\xi$ as function of $T_D$ and $\lambda$. We observe a result that as $T_D$ decreases, $\xi$ increases. It is due to the reason that antipacket dissemination scheme is activated after $T_D$. As a result, if the destination receives the packet earlier, the effects of antipacket dissemination become more prominent, and thus the improvement of fully antipacket dissemination scheme becomes larger. 

      \begin{figure}[t]
          \centering
          \includegraphics[width=5in]{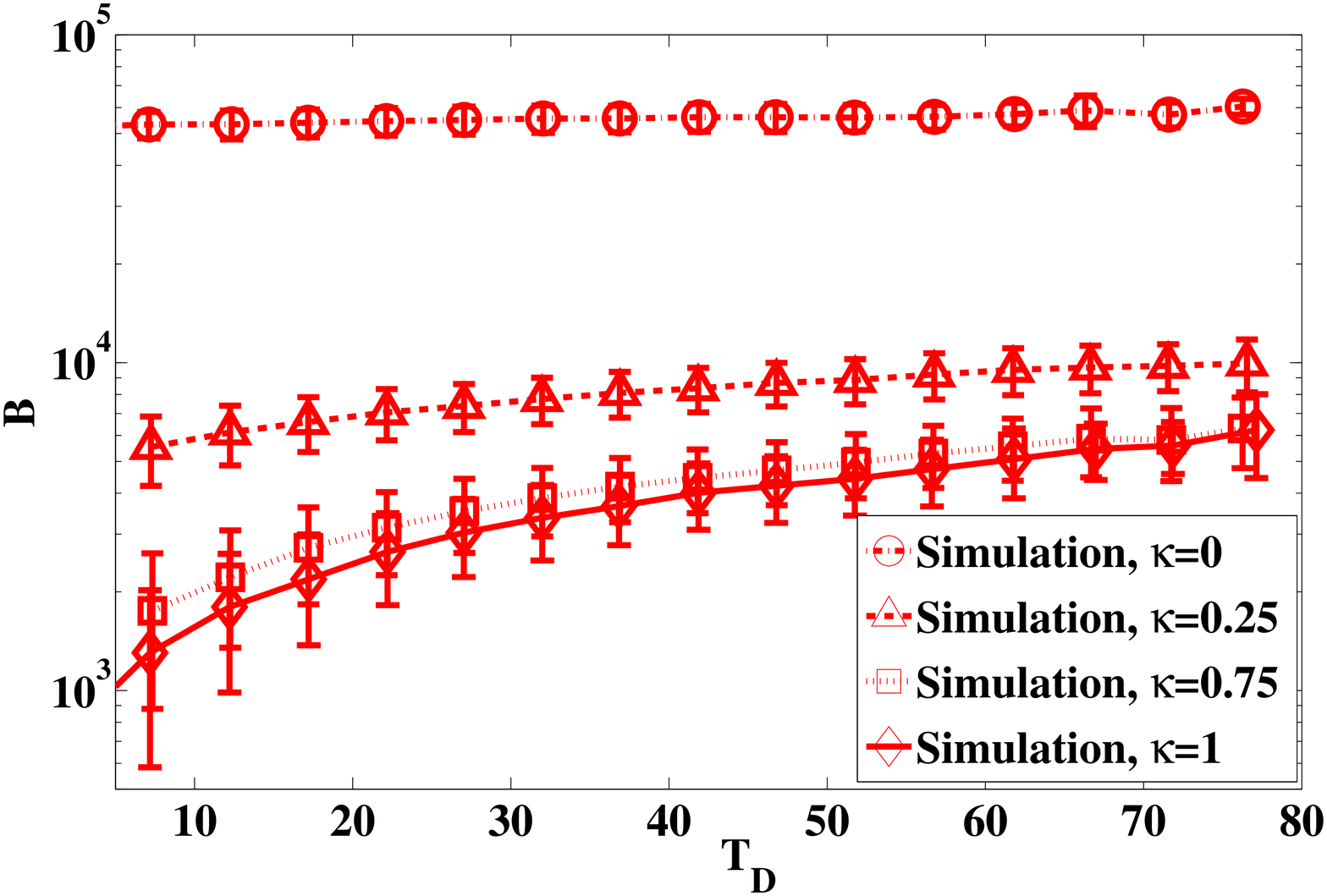}
          \caption{Buffer occupancy with respect to delivery delay $T_D$ and forwarding probability $\kappa$. The system parameters are same as that in Fig.~\ref{fig_anti_k0k1}. The buffer occupancy increases as $\kappa$ decreases or as $T_D$ increases.}
      \label{fig_anti_lambda14_differk_buffer}
      \end{figure}


      \item[Effects of $\kappa$ on $B$.]
      Fig.~\ref{fig_anti_lambda14_differk_buffer} illustrates the effects of $T_D$ and $\kappa$ on buffer occupancy $B$ under fixed pair wise meeting rate $\lambda=0.14817$. As the same trend we found in the previous figures, as $\kappa$ increases, $B$ decreases given the same $T_D$. It is due to the reason that as $\kappa$ becomes larger, the number of nodes participating in antipacket dissemination becomes larger, which facilitates the reduction of buffer occupancy. We can also observe that even with the slight improvement in $\kappa$, the improvement on buffer occupancy is significant, which implies the effectiveness of the antipacket dissemination scheme. 
      \end{LaTeXdescription}

\subsubsection{Comparisons of Different Mobility Models}
      In this subsection we discuss the effect of mobility models on the epidemic routing with antipacket dissemination scheme. Since the correctness of analytical model is validated in the previous subsection, we omit analytical results for the following simulations.

      \begin{figure}[t]
          \centering
          \includegraphics[width=5in]{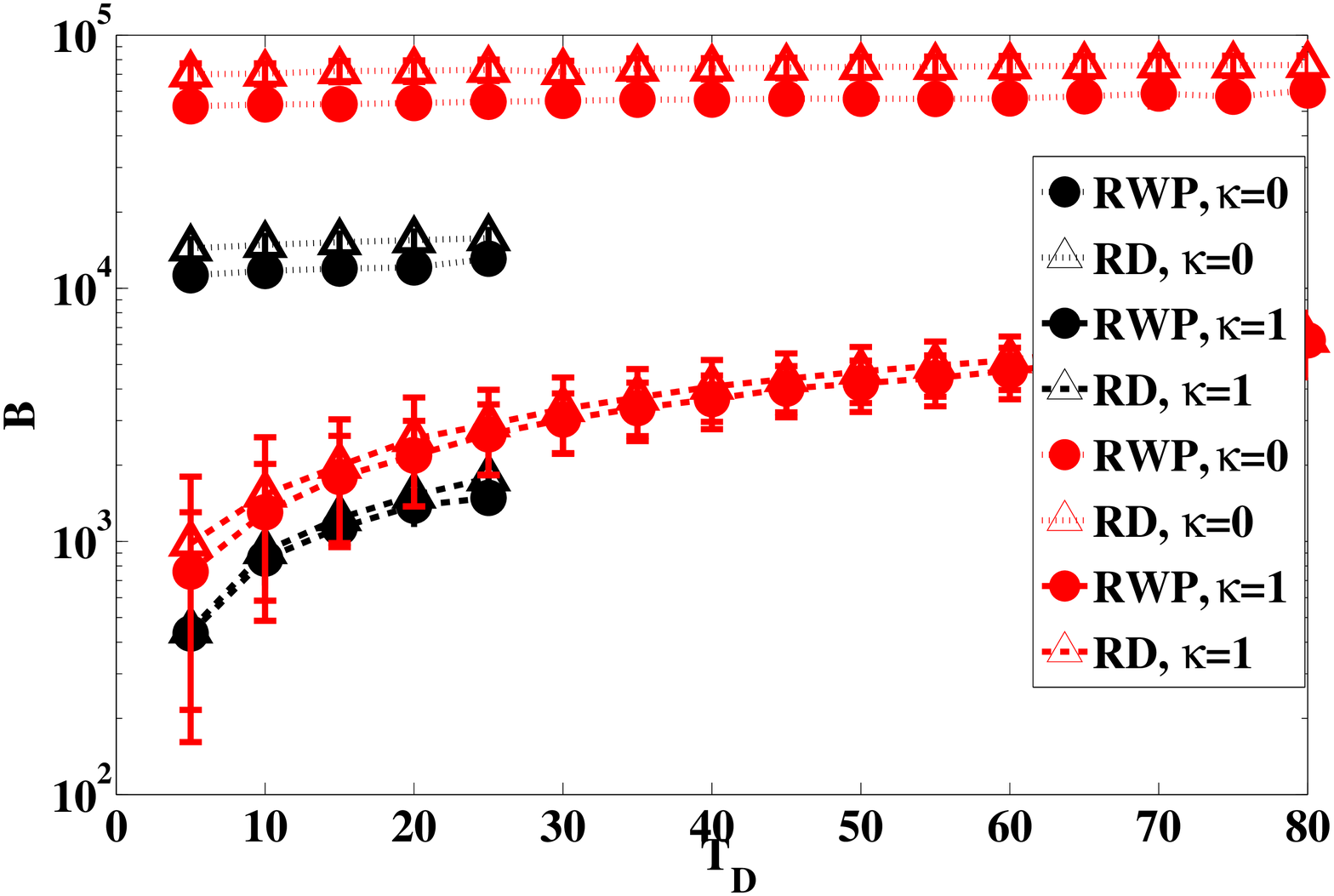}
          \caption{Buffer occupancy with respect to delivery delay with RWP and RD mobility models. The system parameters are same as that in Fig.~\ref{fig_Mobility_TG}. Since RWP performs better than RD, given the same constraint on buffer occupancy, the $\kappa$ has to be set higher in RD than in RWP.}
      \label{fig_Mobility_D_buffer}
      \end{figure}

      \begin{LaTeXdescription}      
      \item[Effects of mobility model on $B$.]
      The effects of $\lambda$ and $T_D$ on $B$ with RD and RWP mobility models are illustrated in Fig.~\ref{fig_Mobility_D_buffer}. Different from what we observed in epidemic routing with global timeout scheme (i.e., Fig.~\ref{fig_Mobility_TG}), RWP performs better than RD in terms of buffer occupancy. It is due to the reason that in antipacket dissemination scheme, both packet and antipacket transmissions rely on the same epidemic paradigm. In particular, if packet spreading is beneficial from a specific mobility model, the antipacket dissemination will be facilitated at the same time. As a result, our observation that RWP is better than RD is consistent with the findings in~\cite{Camp02}. This result also suggests that if the buffer occupancy budget is the same in both mobility schemes, the $\kappa$ have to be set higher in RD than in RWP.
      \end{LaTeXdescription}

      \begin{figure}[t]
          \centering
          \includegraphics[width=5in]{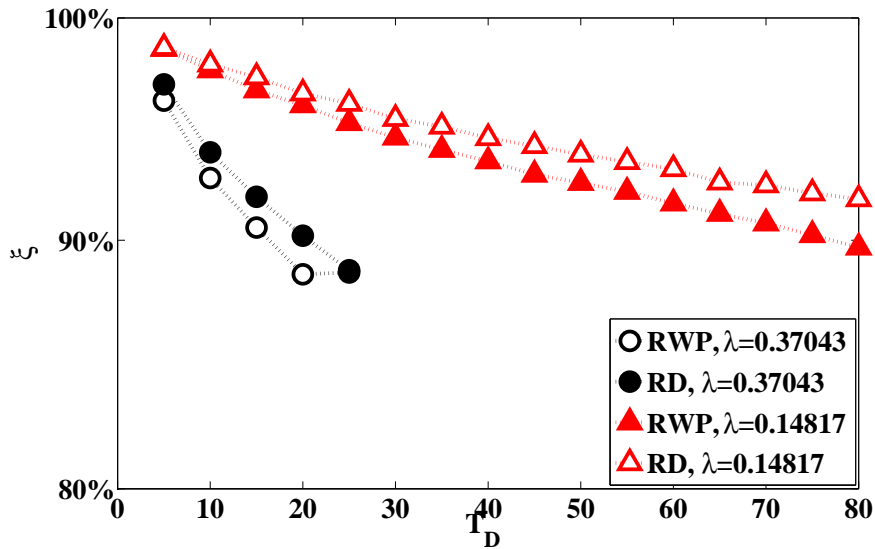}
          \caption{Relative improvement of buffer occupancy with respect to delivery delay and $\lambda$ with RWP and RD mobility models. The system parameters are set as $N=100$, $I_0=1/N$, $T_f=20000$, $r=0.1~km$, $L=2.5352~km$ when $\lambda=0.37043$ and $L=4~km$ when $\lambda=0.14817$, the speed is generated within $4~km/h$ to $10~km/h$. RD is shown to have better relative improvement in buffer occupancy reduction compared to RWP.}
      \label{fig_Mobility_xi}
      \end{figure}

      \begin{LaTeXdescription}      
      \item[Effects of mobility model on $\xi$.]
      Fig.~\ref{fig_Mobility_xi} plots the effects of $\lambda$ and $T_D$ on relative improvement of buffer occupancy $\xi$ with RD and RWP mobility models. It is observed that RD is shown to have better relative improvement in buffer occupancy reduction compared to RWP due to the fact that given the same $T_D$, the buffer occupancy of RD is greater than that of RWP in Fig.~\ref{fig_Mobility_D_buffer}.
   
      \end{LaTeXdescription}

\section{Related Work}
\label{sec_related}
    Epidemic routing is typically applied in intermittently connected mobile network (such as opportunistic network or delay-tolerant network (DTN)) where no permanent end-to-end paths exist between two nodes. The store-and-forward property in epidemic routing achieves successful end-to-end transmission, however, it also incurs extra buffer occupancy to store the replicated packets for forwarding. As a result, how the epidemic routing facilitates the end-to-end transmission becomes the primary topic, and researchers have investigated its performance from the perspectives of delivery delay~\cite{Xin12,Boldrini15,Wang15} or flooding time~\cite{Clementi13,Becchetti14}. Other performance metrics are also receiving attentions, such as packet loss rate~\cite{Socievole11,Clementi13,Preciado14}, transmission cost~\cite{Socievole11,Eshghi14}, infection ratio~\cite{Kim13}, number of copies~\cite{Venkatramanan14} and energy consumption~\cite{Altman13}. Typically, ODEs are exploited to analyze the performance of epidemic routing since ODEs can efficiently capture data dissemination dynamics~\cite{Darling08,CPY11}.  

    Recently, the problem of extra buffer occupancy for epidemic routing has drawn a lot of attentions. Zhang \textit{et al.}~\cite{Zhang07} proposed two approaches, two-hop routing and probabilistic forwarding, to reduce buffer occupancy. In two-hop routing, nodes only forward the message to the destination and the source forwards it to all its neighbors, whereas in probabilistic forwarding, nodes forward the packet to each encountered node with a certain probability. Haas and Small~\cite{Haas06} first proposed immunity schemes for the deletion of unnecessary data packets in epidemic routing. The performance improvement of immunity schemes such as global timeout scheme is evaluated from the aspects of successful transmission probability~\cite{Clementi13,Kim13} and packet loss rate~\cite{Altman13}. De Abreu and Salles~\cite{de2014modeling} further analyzed the lower-bounded value of global timer by estimating the time difference of meetings among nodes, which might not be practical since meeting time is hard to retrieve. 

    Regarding another famous immunity scheme, antipacket dissemination scheme, the effect of (anti)packet on the resource wasting is a critical issue that shall be resolved~\cite{Preciado14}. Eshghi \textit{et al.}~\cite{Eshghi14} introduced a control vector on each node to minimize resource consumption. In our previous work~\cite{CPY11,CPY14}, we combined the global timeout and antipacket dissemination schemes to minimize the buffer occupancy by enabling relay nodes delete the data in a probabilistic fashion upon the expiration of the global timer. To further reduce the unnecessary packets, immunity schemes shall be carefully controlled~\cite{Preciado14,Eshghi14,CPY11,CPY14}. For example, the packet loss rate under a specific value of energy~\cite{Altman13}, the number of copies~\cite{Venkatramanan14}, the number of infected nodes under a specific period~\cite{Kim13} or forwarding policy~\cite{Singh13} are applied to control the immunity scheme. 

    Altman \textit{et al.} determine the optimal probabilistic forwarding policy in order to control the data dissemination \cite{Altman10}. In the later work, they investigate the optimal control policy of two-hop routing with the aid of linear control techniques \cite{Altman11} and separation principle~\cite{Panda15}. Matsuda and Takine \cite{Matsuda08} study the performance of the generalized probabilistic forwarding scheme where each node can relay or discard a packet with certain probability. In \cite{Lin08}, Lin \textit{et al.} use network coding to reduce the buffer occupancy in epidemic routing. Therefore, we can notice that the essence of epidemic routing protocol design is to reduce the buffer occupancy while providing data delivery reliability. However, the optimal control of buffer occupancy is not discussed in both immunity schemes so far. Thus, it still remains open on the tradeoff analysis between buffer occupancy and delivery reliability for epidemic routing.

\section{Conclusion}
\label{sec_con}
      To understand the performance tradeoffs between buffer occupancy and delivery reliability for epidemic routing, we use an SIR model to characterize the state evolution equations of global timeout scheme and antipacket dissemination scheme. For lossy data delivery, we prove the scalability and ubiquity of the global timeout scheme by providing a closed-form expression for optimal global timeout value. With proper selection of the global timeout value as suggested in this paper, the per-node buffer occupancy is shown to only depend on the maximum packet loss rate and pairwise meeting rate, irrespective of the node population, which is crucial for intermittently networking operations. For lossless data delivery, we show that the buffer occupancy can be significantly reduced if every node participates relaying the antipackets to other nodes. End-to-end data transportation is guaranteed while minimizing the buffer occupancy via antipacket dissemination. Consequently, this paper provides performance evaluations and protocol design guidelines for epidemic routing, which offers new insights on buffer occupancy and data delivery reliability analysis. 

\bibliographystyle{IEEEtran}
\bibliography{IEEEabrv,buffer}
\end{document}